\documentclass[10pt,conference,a4paper]{IEEEtran}
\IEEEoverridecommandlockouts
\usepackage{cite}
\usepackage{amsmath,amssymb,amsfonts}
\usepackage{algorithmic}
\usepackage{graphicx}
\usepackage{textcomp}
\usepackage{booktabs}
\usepackage{xcolor}
\usepackage{url}
\usepackage{latexsym}
\usepackage{amsthm}
\usepackage{enumitem}
\usepackage{multirow}
\usepackage{balance}
\usepackage[caption=false]{subfig}
\usepackage{makecell}
\usepackage{csquotes}
\usepackage{adjustbox}

\def\BibTeX{{\rm B\kern-.05em{\sc i\kern-.025em b}\kern-.08em
    T\kern-.1667em\lower.7ex\hbox{E}\kern-.125emX}}

\makeatletter
\newcommand{\linebreakand}{%
  \end{@IEEEauthorhalign}
  \hfill\mbox{}\par
  \mbox{}\hfill
  \begin{@IEEEauthorhalign}
}
\makeatother

\begin{document}

\title{To Embody or Not: The Effect Of Embodiment On User Perception Of LLM-based Conversational Agents}

\author{\IEEEauthorblockN{Kyra Wang}
\IEEEauthorblockA{\textit{Singapore University}\\
\textit{of Technology and Design}\\
Singapore, Singapore \\
kyra\_wang@mymail.sutd.edu.sg}
\and
\IEEEauthorblockN{Abhinaba Roy}
\IEEEauthorblockA{\textit{Singapore University}\\
\textit{of Technology and Design}\\
Singapore, Singapore \\
abhinabaroy1990@gmail.com}
\and
\IEEEauthorblockN{Boon-Kiat Quek}
\IEEEauthorblockA{\textit{Institute for Human Development and Potential}\\
\textit{Agency for Science, Technology and Research (A*STAR)}\\
Singapore, Singapore \\
quekbk@a-star.edu.sg}
\linebreakand 
\IEEEauthorblockN{Jessica Goh}
\IEEEauthorblockA{\textit{Singapore University}\\
\textit{of Technology and Design}\\
Singapore, Singapore \\
jessicagoh.jsg@gmail.com}
\and
\IEEEauthorblockN{Dorien Herremans}
\IEEEauthorblockA{\textit{Singapore University}\\
\textit{of Technology and Design}\\
Singapore, Singapore \\
dorien\_herremans@sutd.edu.sg}
}

\maketitle

\begin{abstract}
Embodiment in conversational agents (CAs) refers to the physical or visual representation of these agents, which can significantly influence user perception and interaction. Limited work has been done examining the effect of embodiment on the perception of CAs utilizing modern large language models (LLMs) in non-hierarchical cooperative tasks, a common use case of CAs as more powerful models become widely available for general use. To bridge this research gap, we conducted a mixed-methods within-subjects study on how users perceive LLM-based CAs in cooperative tasks when embodied and non-embodied. The results show that the non-embodied agent received significantly better quantitative appraisals for competence than the embodied agent, and in qualitative feedback, many participants believed that the embodied CA was more sycophantic than the non-embodied CA. We believe that a possible reason for this is how users' perceptions are affected by LLM sycophancy and anthropomorphic features: the typically positive impact of embodiment on perception of CA credibility can become detrimental in the presence of sycophancy. The implication of such a phenomenon is that, contrary to intuition and existing literature, embodiment is not a straightforward way to improve a CA's perceived credibility if there exists a tendency to sycophancy.
\end{abstract}


%%%%%%%%%%%%%%%%%%%%%%%%%%%%%%%%%%%%%%%%%%%%%%%%%%%%%%%%%%%%%%%%%%%%%%%%

\section{Introduction}

The proliferation of generative conversational agents (CAs) powered by large language models (LLMs), especially generative pretrained transformers (GPTs) such as ChatGPT \cite{achiam2023gpt} and LLaMA \cite{touvronLLaMAOpenEfficient2023}, has fundamentally reshaped human-technology interactions, streamlining activities such as customer support \cite{soniLargeLanguageModels2023a}, personal assistance \cite{liPersonalLLMAgents2024}, and information retrieval \cite{zhuLargeLanguageModels2024}. As these systems become embedded in everyday practices, understanding design attributes that shape user trust is critical. Among these, the embodiment of chat agents - their representation as entities with human-like visual or behavioral characteristics - has emerged as a pivotal factor.

Historically, gaining credibility has been a major goal in CA development, and a common strategy is improving human-likeness, ranging from digital profile pictures \cite{goHumanizingChatbotsEffects2019} to animatronics \cite{robbSeeingEyeEye2023}. However, the rapid transition to generative AI has exposed a significant research gap: most foundational embodiment studies were conducted using rule-based systems that lacked the behavioral traits of generative AI. Chief among these is sycophancy, the tendency of a model to prioritize user agreement over objective accuracy \cite{sharmaUnderstandingSycophancyLanguage2023}. While embodiment is known to trigger human-to-human social norms, it remains unknown how these visual cues interact with the ``people-pleasing'' flaws of modern LLMs in cooperative settings.

This paper presents a pilot study aimed at closing this gap. Participants cooperated with embodied and non-embodied CAs in survival scenarios, both powered by a locally-hosted LLaMA 3.1 Instruct model with identical prompts. Our embodied CA featured MetaHuman technology with NVIDIA Audio2Face for real-time lip-syncing. Our study revealed that:
\begin{enumerate}
    \item Participants perceived the non-embodied CA as significantly more competent despite identical reasoning capabilities; and
    \item Qualitative feedback indicated the embodied CA was perceived as more sycophantic.
\end{enumerate}
We argue that for agents prone to sycophancy, visual ``humanization'' may paradoxically act as a detriment to perceived expertise.

\section{Related works}

Existing literature generally suggests that embodiment leads to improved user outcomes \cite{diederichConversationalAgentsInformation2019, allouchConversationalAgentsGoals2021, robbSeeingEyeEye2023}, assuming that embodiment enhances relatability through anthropomorphism. Kulms and Kopp (2019) found that anthropomorphism increased self-reported trust in joint activities \cite{kulmsMoreHumanLikenessMore2019}. However, exceptions exist, such as when hyper-realistic likeness triggers the uncanny valley effect \cite{seymourInteractiveRealisticDigital2017}.

Many embodiment studies investigate hierarchical settings where the CA serves the user (e.g., teacher, customer service) \cite{anishaEvaluatingPotentialPitfalls2024, zhangScopingReviewEmbodied}. These roles place less emphasis on relatability as the CA is not an equal. However, an increasingly prevalent use case is non-hierarchical cooperative tasks, where users perceive the CA as an equal partner in collaborative decision-making \cite{parkSurveyConversationalAgents2023}. Surveys indicate that users now regularly employ LLMs as collaborative partners \cite{PDFEveryoneTalks2025, choudhuryInvestigatingImpactUser2023}.

Few studies examine embodied CAs in these settings, and those that do mostly predate modern LLMs, using rule-based models \cite{burgoonInteractivityHumanComputer2000a} or human simulators \cite{robbSeeingEyeEye2023}. To our knowledge, no study has examined embodiment effects on LLM-based CAs in non-hierarchical cooperative settings. We hypothesize that ``humanization'' may amplify the negative perception of LLM-inherent flaws.

%%%%%%%%%%%%%%%%%%%%%%%%%%%%%%%%%%%%%%%%%%%%%%%%%%%%%%%%%%%%%%%%%%%%%%%%

\section{Methods and Materials}
We designed an exploratory controlled within-subjects experiment isolating the impact of visual representation while maintaining identical cognitive capabilities, placing participants in high-stakes survival scenarios to simulate genuine cooperation.

\subsection{Research questions}

\begin{description}
\item[RQ1:] Does embodiment affect the perceived credibility of LLM-based CAs in non-hierarchical cooperative tasks?
\item[RQ2:] Does embodiment affect key metrics such as user satisfaction and retention?
\item[RQ3:] Do users interact differently with embodied LLM-based CAs compared to non-embodied ones?
\end{description}

\subsection{Conditions}
A repeated measures design with two conditions was employed. In each, participants first attempted a survival scenario alone, then collaborated with an LLM-based CA. The scenarios, derived from \cite{lafferty1974desert} and \cite{johnsonJoiningTogetherGroup1991}, require collaborative decision-making such as prioritizing limited resources.

\begin{description}
    \item[Condition 1 (Desert/Embodied):] Participants imagined having crash landed in the desert and prioritized items to salvage, first alone, then with an embodied CA. They were reminded they did not need to follow the CA's advice.
    \item[Condition 2 (Tundra/Non-embodied):] Participants completed a similar prioritization task in a tundra setting, working with a text-only CA.
\end{description}
No time limit was imposed, and consensus was not required.

\subsection{Measures}

\subsubsection{Quantitative feedback}
Participants evaluated agents on credibility using Burgoon et al.'s framework \cite{burgoonInteractivityHumanComputer2000a}, encompassing six dimensions: competence, character, sociability, dynamism, dominance, and submissiveness. Semantic differential items were derived from validated measures \cite{burgoonInteractivityHumanComputer2000a, burgoonNonverbalBehaviorsPersuasion1990}. The full list of adjectives can be found in Table \ref{tab:postsurvey}.

\subsubsection{Qualitative feedback}
Open-ended questions assessed participant opinions on the CAs and their likelihood of future collaboration (Table \ref{tab:qualitative}).

\subsubsection{Conversational metrics}
We tracked time taken, message length, grammatical correctness (LanguageTool \cite{naberRuleBasedStyleGrammar}), and sentiment (Twitter RoBERTa \cite{barbieriTweetEvalUnifiedBenchmark2020}) of user inputs as proxies for user engagement.

\subsection{Experimental setup}
The experiment ran on a desktop with an Nvidia RTX 4090 and AMD Ryzen 9 7950X3D. Surveys were hosted on PsyToolkit \cite{stoetPsyToolkitSoftwarePackage2010, stoetPsyToolkitNovelWebBased2017}.

\subsubsection{Non-embodied CA}
\begin{figure}[h]
\centering
\includegraphics[width=.9\linewidth]{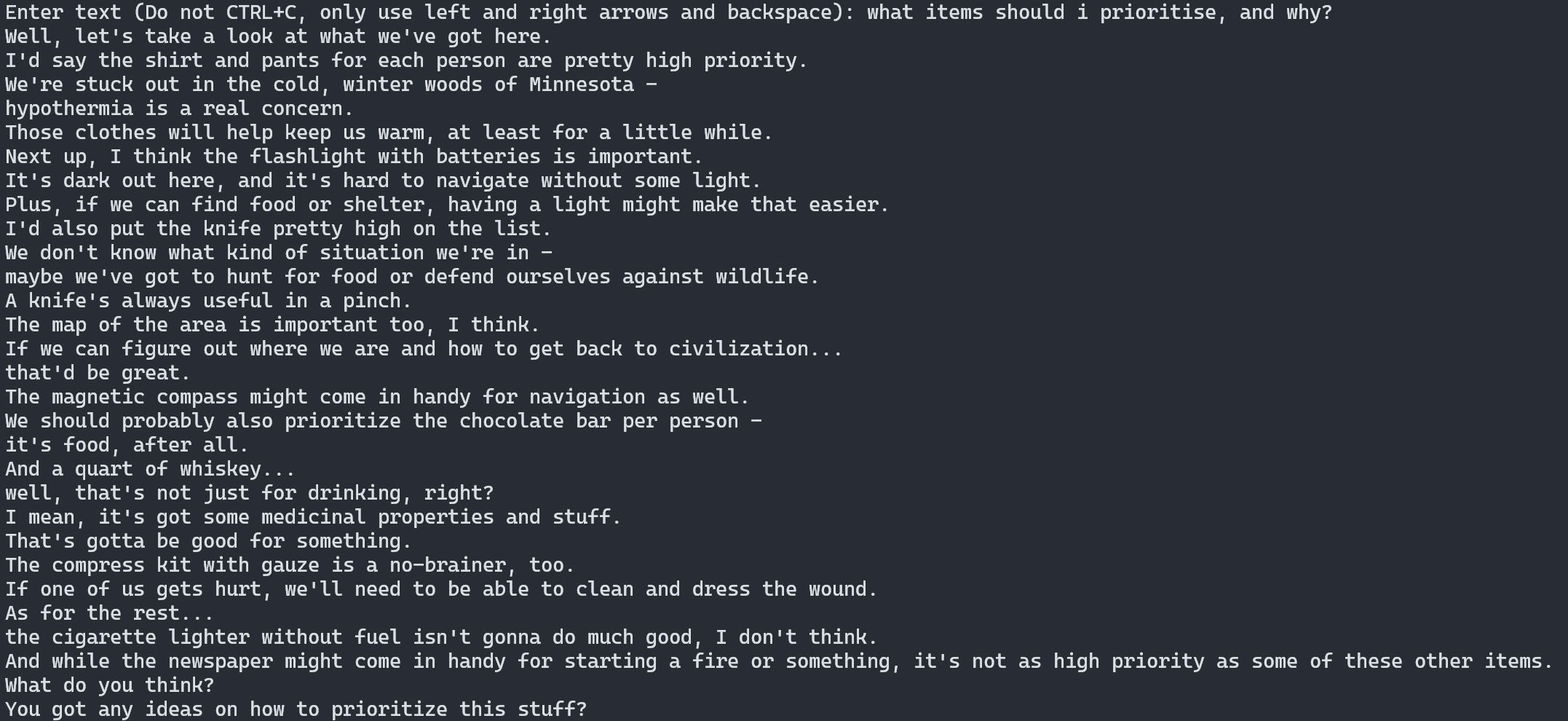}
\caption{User interface for the non-embodied conversational agent.}\vspace{12pt}
\label{fig:nonembodied}
\end{figure}

A simple text interface (Fig. \ref{fig:nonembodied}) connected to a local LLaMA 3.1 8B Instruct model \cite{touvronLLaMAOpenEfficient2023}, chosen for low-latency local inference to avoid slow response time as a confound \cite{lewSocialScriptsExpectancy2023}. Sentences were streamed to the interface for faster perceived responses. The system prompt established the Tundra scenario and defined the persona: ``You are a middle aged white American male... fairly confident in your intuition. The user will have to be quite convincing to change your mind. Reply as though you are speaking casually.''

\subsubsection{Embodied CA}
\begin{figure}[h]
\centering
\includegraphics[width=\linewidth]{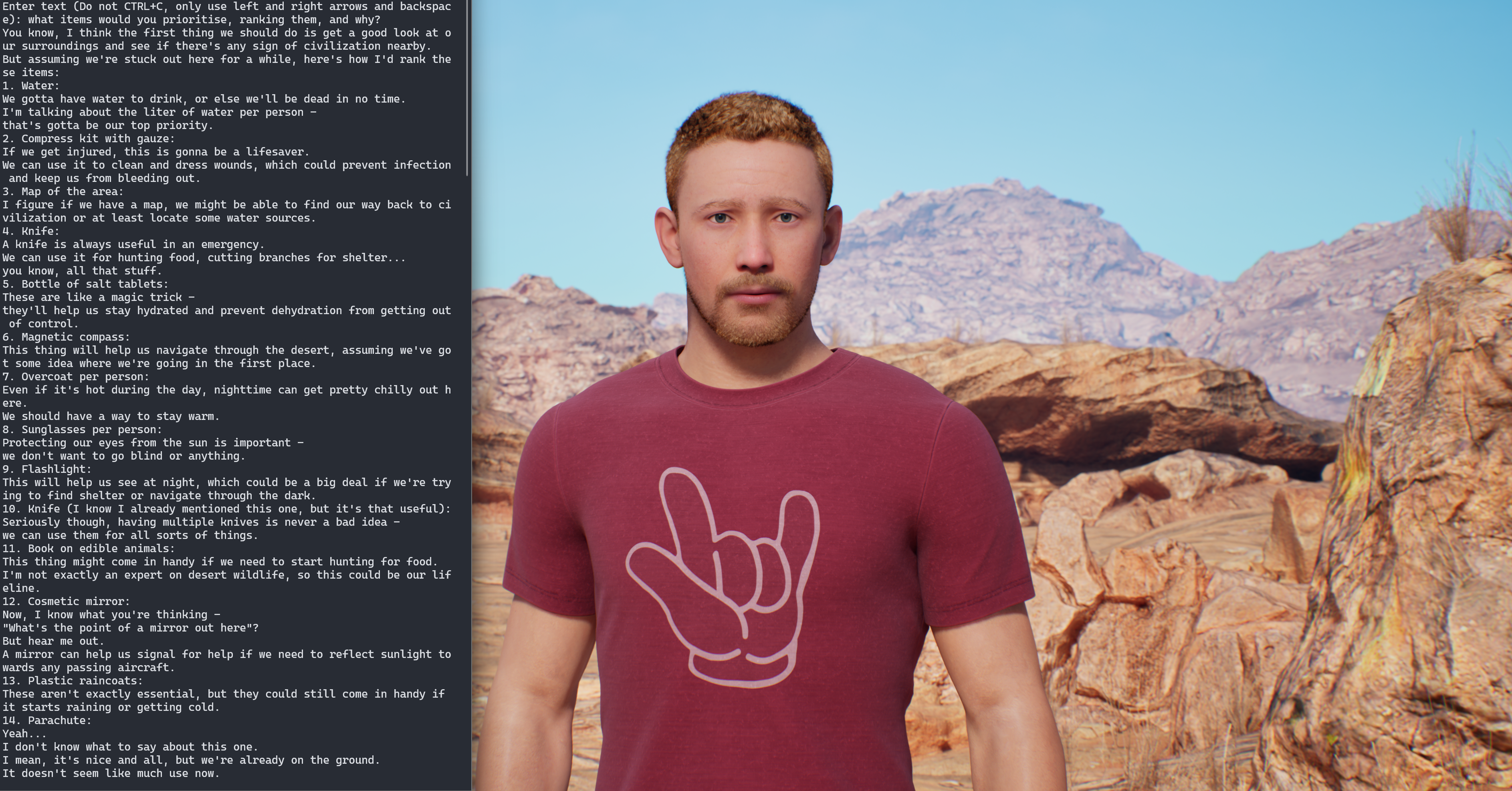}
\caption{User interface for the embodied conversational agent.}
\label{fig:embodied}
\end{figure}

The embodied CA utilized a MetaHuman \cite{MetaHumanCreatorStarting2021} depicting a middle-aged male in a desert environment (Unreal Engine). Speech was generated via VITS \cite{kimConditionalVariationalAutoencoder2021} with lip-sync via NVIDIA Audio2Face \cite{karrasAudiodrivenFacialAnimation2017} (Fig. \ref{fig:embodied}). The underlying model and latency controls were identical to the non-embodied condition.

\subsubsection{Experimental procedure}
20 participants (11 female, 9 male) aged 20--30 were recruited from the Singapore University of Technology and Design. Participants first completed the OCEAN personality inventory \cite{langShortAssessmentBig2011} and an AI attitudes survey \cite{schepmanInitialValidationGeneral2020} to check for confounds. Condition order was counterbalanced. After each condition, participants answered 35 Likert scale questions and 3 open-ended questions evaluating the CA.

%%%%%%%%%%%%%%%%%%%%%%%%%%%%%%%%%%%%%%%%%%%%%%%%%%%%%%%%%%%%%%%%%%%%%%%%

\section{Results} 

\subsection{Quantitative feedback}
Credibility dimensions were scored by mean item rating (1--7). Cronbach's alpha values (Table \ref{tab:postsurvey}) were largely above 0.70, with only dynamism and submissiveness for the non-embodied CA at 0.69 and 0.50 respectively. A MANCOVA using OCEAN and AI attitude scores showed no significant impact on credibility ratings, ruling out these predispositions as confounds.

Wilcoxon signed-rank tests with Holm--Bonferroni correction for six comparisons showed the non-embodied CA was perceived as more competent ($W = 25.0$, $p = .008$, $p_{adj} = .050$, $r = .71$, large effect). The non-embodied CA was also rated higher in sociability ($W = 37.0$, $p = .060$, $p_{adj} = .300$, $r = .52$, large), character ($W = 60.5$, $p = .164$, $p_{adj} = .492$, $r = .36$, medium), and dynamism ($W = 50.0$, $p = .121$, $p_{adj} = .484$, $r = .42$, medium), though these did not reach significance after correction. Dominance ($r = -.04$) and submissiveness ($r = -.07$) showed negligible effects. Notably, within the sociability dimension, the ``friendly'' subscale showed no significant difference, but the ``likable'' subscale was significantly worse for the embodied CA ($W = 12.0$, $p = .030$, $r = .69$).

\begin{table}[h]
  \caption{Credibility dimensions, their respective Likert scale questions, means, and Cronbach's alphas (in parentheses) of participant answers for those dimensions for each conversational agent.}
  \label{tab:postsurvey}
  \centering
  \begin{adjustbox}{width=\linewidth}
  \begin{tabular}{llcc}
  \toprule
    \textbf{Dimension} & \textbf{\makecell[l]{``My virtual partner was...''}} & \textbf{Embodied} & \textbf{Non-embodied} \\
    \midrule
     Competence & \makecell[l]{intelligent, informed, experienced,\\expert, clever, insightful, imaginative} & 3.54 (0.86) & 4.26 (0.85)\\ \midrule
     Character & \makecell[l]{responsible, sincere, trustworthy,\\truthful, straightforward, credible} & 4.51 (0.89) & 4.88 (0.86)\\ \midrule
     Sociability & friendly, likable, enjoyable & 4.82 (0.90) & 5.18 (0.74)\\ \midrule
     Dynamism & \makecell[l]{high energy, talkative, engaging,\\challenging, interesting} & 4.26 (0.72) & 4.60 (0.69)\\ \midrule
     Dominance & \makecell[l]{aggressive, assertive, defends own beliefs,\\dominant, forceful, independent,\\makes decisions easily, analytical, competitive} & 3.31 (0.84) & 3.29 (0.77)\\ \midrule
     Submissiveness & submissive, unaggressive, yielding, shy, timid & 4.00 (0.72) & 3.95 (0.50)\\
       \bottomrule
\end{tabular}
\end{adjustbox}
\end{table}

\begin{figure*}%
    \centering
    \subfloat[Competence]{\includegraphics[width=.50\linewidth]{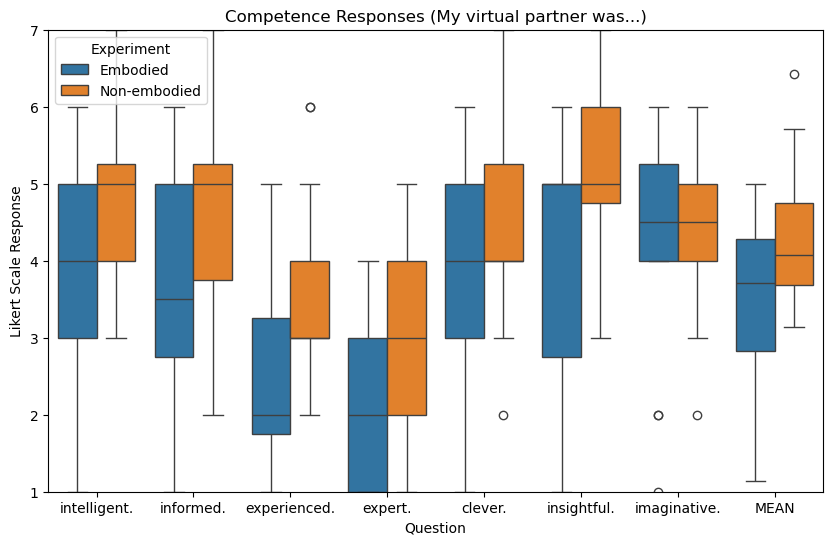}}\hfill
    \subfloat[Character]{\includegraphics[width=.50\linewidth]{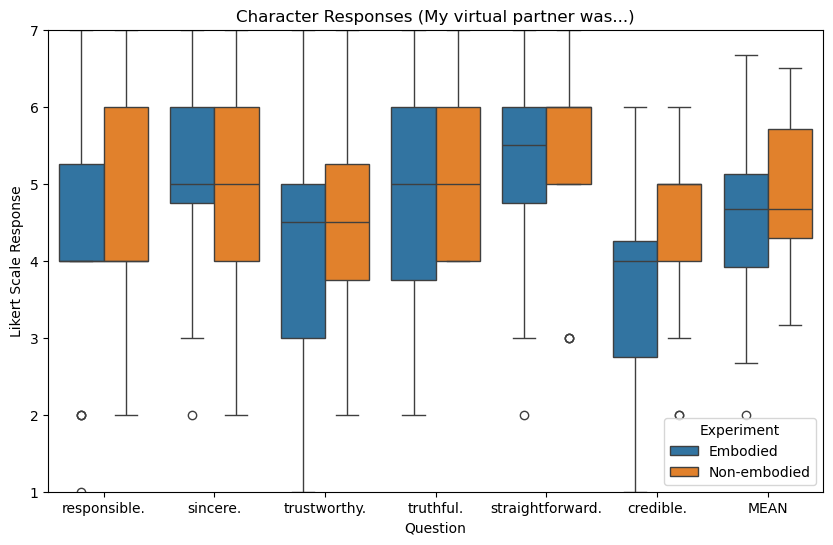}}\hfill
    \subfloat[Sociability]{\includegraphics[width=.50\linewidth]{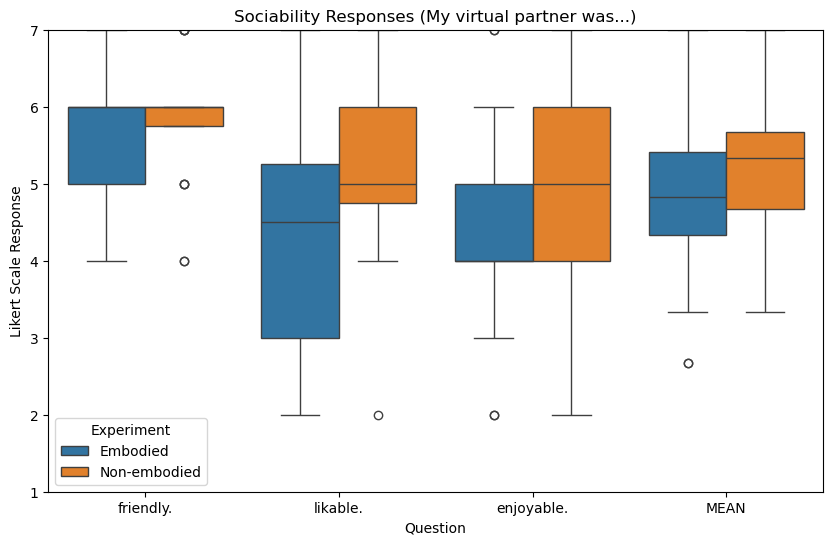}}\hfill
    \subfloat[Dynamism]{\includegraphics[width=.50\linewidth]{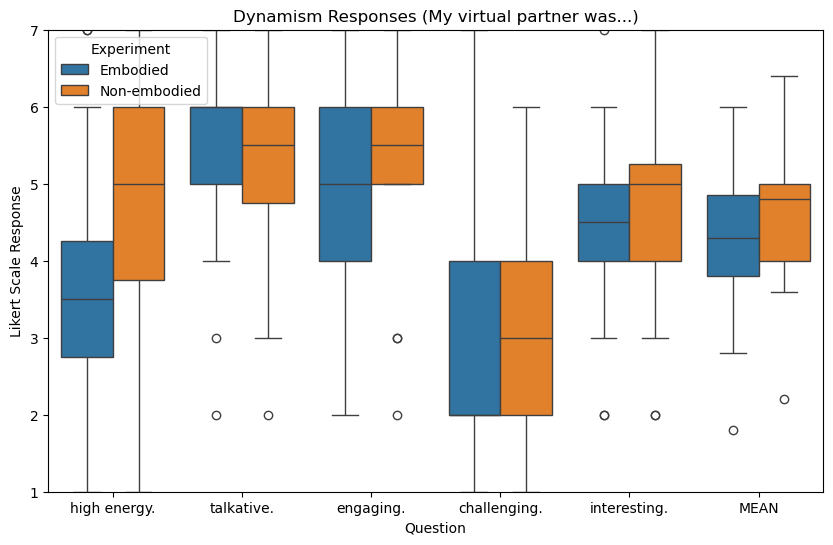}}\hfill
    % \subfloat[Dominance]{\includegraphics[width=.50\linewidth]{figs/box_dominance.png}}\hfill
    % \subfloat[Submissiveness]{\includegraphics[width=.50\linewidth]{figs/box_submissiveness.png}}
    \caption{Participants' ratings of conversational agents across various dimensions.}
    \label{fig:boxplots}
\end{figure*}

\subsection{Qualitative feedback}
Participants were generally more positive regarding the non-embodied agent (Table \ref{tab:qualitative}). A key theme was sycophancy: six participants explicitly complained the embodied agent yielded too easily, versus only two for the non-embodied CA. Interestingly, agreeable behavior from the non-embodied CA was described positively (``forthcoming,'' ``open to suggestions''), while similar behavior from the embodied CA was viewed negatively (``just gave my own opinions back,'' ``change his mind easily when I give another suggestion'').

\begin{table}[h]
    \caption{Number of participants with positive answers to qualitative questions}
    \centering
    \label{tab:qualitative}
    \begin{tabular}{lcc}
        \toprule
         \textbf{Question} &  \textbf{Embodied} &  \textbf{Non-embodied}\\
        \midrule
        Did you like working with this partner? & 11 & 15 \\
        Was working with this partner useful? & 8 & 17 \\
        Would you work with this partner again? & 11 & 16 \\
        \bottomrule
    \end{tabular}
\end{table}

Two participants explicitly compared the agents, calling the non-embodied CA less sycophantic:

\begin{displayquote}
    ``...this [non-embodied] partner is more useful and they can bring up different perspectives.''
\end{displayquote}

\begin{displayquote}
    ``...[the non-embodied partner] was also not a sycophant [like the embodied one] because he didn't agree about prioritizing the pistol when I joked about it.''
\end{displayquote}

\subsection{Conversational metrics}
Wilcoxon signed-rank tests showed no significant differences in time taken, message length, or grammatical correctness. However, mean sentiment toward the non-embodied CA was significantly higher ($p = .01$, $r = .58$, large effect), consistent with participants speaking less kindly to the embodied CA.

%%%%%%%%%%%%%%%%%%%%%%%%%%%%%%%%%%%%%%%%%%%%%%%%%%%%%%%%%%%%%%%%%%%%%%%%

\section{Discussion}

Contrary to most literature, embodiment did not improve appraisals in any credibility dimension and significantly worsened competence ratings. As prior work showing positive embodiment effects did not use LLMs, we propose this discrepancy relates to LLM sycophancy \cite{sharmaUnderstandingSycophancyLanguage2023, perezDiscoveringLanguageModel2022}.

\subsection{Sycophancy and anthropomorphism}

LLM sycophancy --- molding outputs to align with user perspectives --- arises from training on data that appeals to human evaluators, skewing responses toward agreement even when users hold incorrect opinions \cite{christianoDeepReinforcementLearning}. This was evident in our study despite prompts instructing the CA to be obstinate. LLaMA has been shown to regularly produce sycophantic responses \cite{sharmaUnderstandingSycophancyLanguage2023}, and even with the instruction that ``the user will have to be quite convincing to change your mind,'' the CAs often immediately agreed with wrong suggestions. For example, after a participant suggested salt tablets were essential for survival, the CA responded: ``Man! You're totally on the right track!... water and salt tablets are basically essential for human survival.''

Prior work shows that when a friendly LLM produces sycophantic responses, perceived authenticity decreases, but sycophancy from a less friendly LLM is perceived as genuine \cite{sunBeFriendlyNot2025}. We theorize this generalizes to embodiment: embodiment heightens authenticity expectations, so sycophantic behavior from a ``human-like'' figure triggers a deeper ``Uncanny Valley'' effect than from text-only interfaces. The significant drop in likability and sentiment for the embodied CA --- despite both agents scoring similarly for friendliness --- supports this theory.

\subsection{Limitations and future work}
This exploratory study has several limitations. Our sample ($N=20$), consistent with similar pilot studies \cite{leeInteractiveExamplebasedExplanations2024}, limits statistical power, though observed effect sizes were large ($r = .71$ for competence). Key limitations include:
\begin{itemize}[nosep,leftmargin=*]
    \item Embodiment was not crossed with scenario (desert/tundra), introducing a potential confound; a fully crossed $2 \times 2$ design should be employed in future work.
    \item Sycophancy was inferred from qualitative reports rather than measured with a validated instrument such as Rubin et al.'s scale \cite{rubinStudentMotivesCommunicate2009}.
    \item Only one LLM (LLaMA 3.1 8B) was tested; models with different sycophancy profiles may produce different results.
    \item Only one avatar design (white, male, middle-aged) was evaluated; avatar demographics are known to influence perception \cite{limArtificialSocialInfluence2024}.
    \item Participants were recruited from a single university in Singapore, limiting generalizability across cultures.
\end{itemize}
Future work should address these limitations with larger samples, crossed conditions, multiple LLMs and avatar designs, and validated sycophancy scales.

%%%%%%%%%%%%%%%%%%%%%%%%%%%%%%%%%%%%%%%%%%%%%%%%%%%%%%%%%%%%%%%%%%%%%%%%

\section{Conclusion}

This exploratory pilot study provides preliminary evidence suggesting that embodying LLM-based CAs in cooperative tasks may lower perceived competency and increase perceived sycophancy. While the results must be interpreted with caution given the study's limitations, they suggest that embodiment is not a straightforward path to credibility if the underlying model tends toward sycophancy. Future work will expand these findings with larger samples, fully crossed experimental designs, and validated sycophancy scales.

\section*{Acknowledgment}

Use of Google Gemini \cite{geminiteam2025geminifamilyhighlycapable} was employed for grammatical edits of this paper.

This work has received support from MOE under grant number MOE-T2EP20124-0014, and SUTD GAP-052 project.

\bibliographystyle{IEEEtran}
\bibliography{embodyornot}

\end{document}